\documentclass[aps,reprint]{revtex4-2}

\usepackage{graphicx}
\usepackage{mathtools}
\usepackage{amsmath}
\usepackage{amssymb}
\usepackage{newtxmath}
\usepackage{subcaption}
\usepackage{chemfig}
\usepackage{float}
\usepackage{calc}
\usepackage{pgfplotstable}
\usepackage[normalem]{ulem}
\usepackage{todonotes}
\usepackage{adjustbox}
\usepackage[hidelinks]{hyperref}
\usepgfplotslibrary{groupplots}
\usetikzlibrary{backgrounds}

\definecolor{plot1}{HTML}{f3622d}
\definecolor{plot2}{HTML}{fba71b}
\definecolor{plot3}{HTML}{57b757}
\definecolor{plot4}{HTML}{41a9c9}
\definecolor{plot5}{HTML}{4258c9}
\definecolor{plot6}{HTML}{9a42c8}
\definecolor{plot7}{HTML}{c84164}
\definecolor{plot8}{HTML}{888888}

\newcommand{\crule}[3][black]{\textcolor{#1}{\rule{#2}{#3}}}
\newcommand{\av}[1]{\left< #1 \right>}

\begin{document}

\preprint{Pre-Print}

\title{Revealing contributions to conduction from transport within ordered and disordered regions in highly doped conjugated polymers through analysis of temperature-dependent Hall measurements}

\author{William A. Wood}
\email{waw31@cam.ac.uk}
\author{Ian E. Jacobs}
\author{Leszek J. Spalek}
\author{Yuxuan Huang}
\author{Chen Chen}
\author{Xinglong Ren}
\author{Henning Sirringhaus}
\email{hs220@cam.ac.uk}
\affiliation{Cavendish Laboratory, University of Cambridge}

\date{\today}

\begin{abstract}
    Hall effect measurements in doped polymer semiconductors are widely
    reported, but are difficult to interpret due to screening of Hall voltages
    by carriers undergoing incoherent transport. Here, we propose a refined
    analysis for such Hall measurements, based on measuring the Hall coefficient
    as a function of temperature, and modelling carriers as existing in a regime
    of variable ``deflectability'' (i.e. how strongly they ``feel'' the magnetic
    part of the Lorentz force). By linearly interpolating each carrier between
    the extremes of no deflection and full deflection, we demonstrate that it is
    possible to extract the (time-averaged) concentration of deflectable charge
    carriers, $\av{n_d}$, the average, temperature-dependent mobility of those
    carriers, $\av{\mu_d}(T)$, as well as the ratio of conductivity that comes
    from such deflectable transport, $d(T)$. Our method was enabled by the
    construction of an improved AC Hall measurement system, as well as an
    improved data extraction method. We measured Hall bar devices of
    ion-exchange doped films of PBTTT-C$_{14}$ from 10\,--\,300\,K. Our analysis
    provides evidence for the proportion of conductivity arising from
    deflectable transport, $d(T)$, increasing with doping level, ranging between
    15.4\% and 16.4\% at room temperature. When compared to total
    charge-carrier-density estimates from independent methods, the values of
    $\av{n_d}$ extracted suggest that carriers spend $\sim$\,37\% of their time
    of flight being deflectable in the most highly doped of the devices measured
    here. The extracted values of $d(T)$ being less than half this value thus
    suggest that the limiting factor for conductivity in such highly doped
    devices is carrier mobility, rather than concentration.
\end{abstract}

\maketitle

\section{Introduction}

\begin{figure*}[htb]

    \centering
    \begin{subfigure}[t]{0.64\linewidth}
        \caption{}
        \centering
        \includegraphics{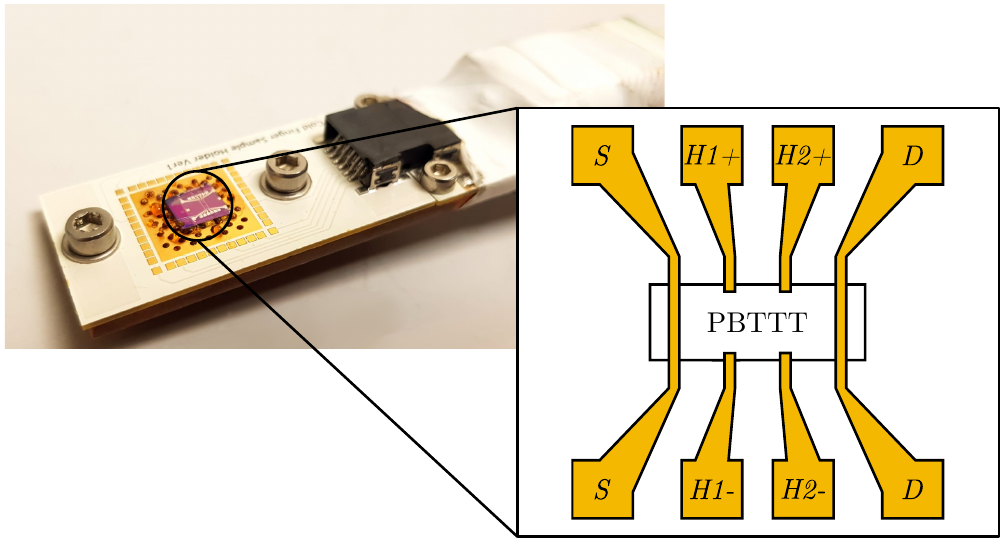}
        \label{fig:first:device}
    \end{subfigure}
    \hfill
    \begin{subfigure}[t]{0.32\linewidth}
        \caption{}
        \centering
        \includegraphics[width=\linewidth]{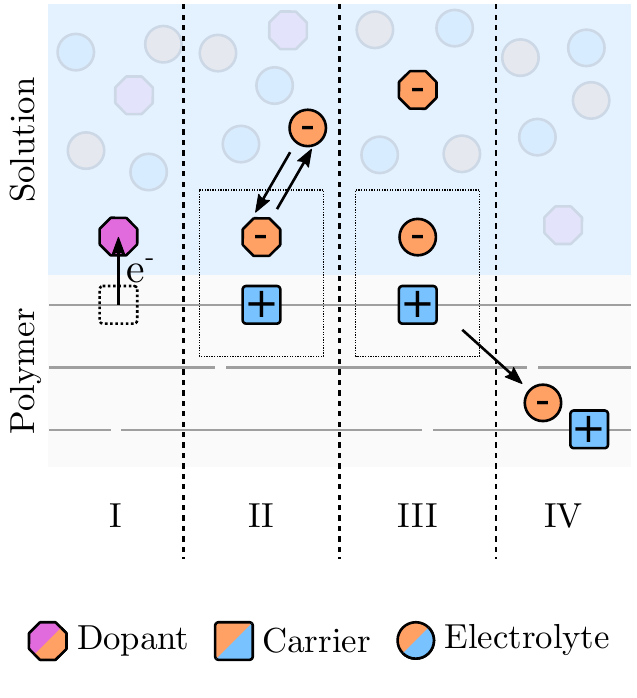}
        \label{fig:first:iex}
    \end{subfigure}\\
    \begin{subfigure}[t]{\linewidth}
        \caption{}
        \centering
        \includegraphics[width=\linewidth]{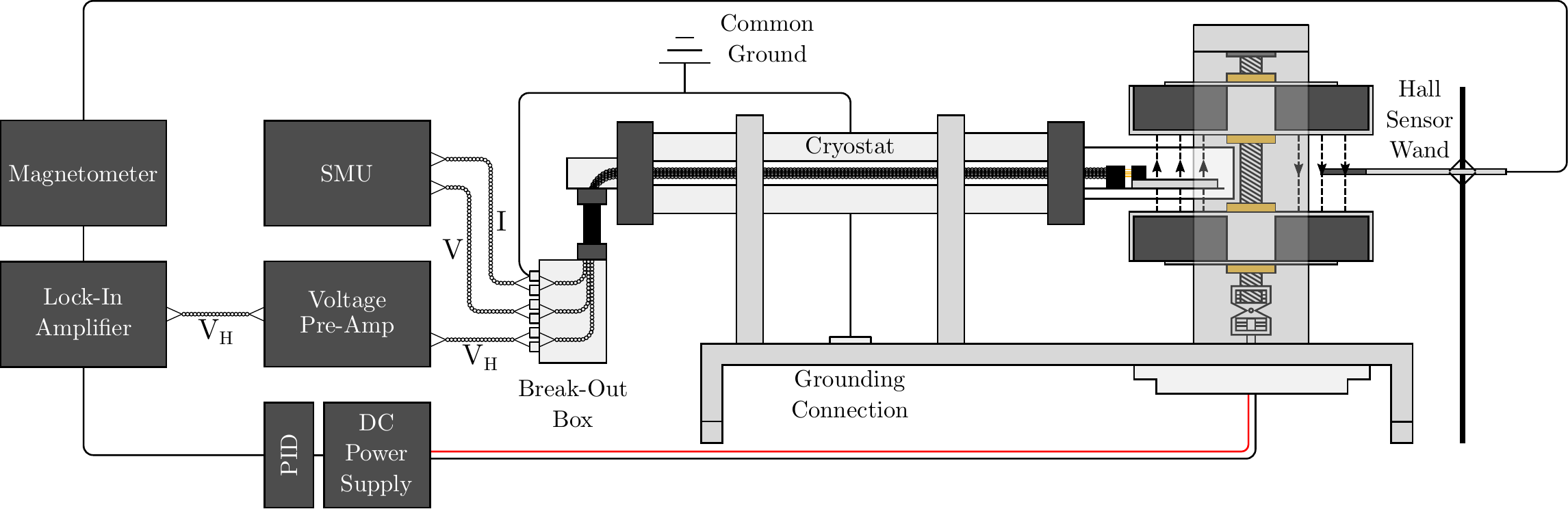}
        \label{fig:first:setup}
    \end{subfigure}

    \caption{PBTTT Hall measurements: (\subref{fig:first:device}) PBTTT Hall bar
    loaded on cryostat via custom-made sample board with device schematic
    displayed; (\subref{fig:first:iex}) Diagram displaying the stages of the
    ion-exchange doping process (I: charge transfer, II to III: exchange of
    ions, IV: charge carrier and exchanged ion diffuse into polymer);
    (\subref{fig:first:setup}) Diagrammatic representation of the AC Hall
    measurement system used for measurements.}
    \label{fig:first}
\end{figure*}

Since the original, Nobel-prize-winning discovery of substantial electrical
conductivities in polyacetylene by Heeger et al,\cite{Heeger1977} there has been
continued interest in understanding the charge-transport physics of such
degenerately doped, conjugated polymers. Recent advances in doping methods have
focused on minimising structural disorder associated with incorporating dopant
ions into polymer films, thus reaching optimal electrical conductivities. An
example of this is the recently explored technique of ion-exchange (IEx) doping,
where a polymer film is exposed to a solution containing both a molecular dopant
and an electrolyte. After electron transfer from the polymer, the ionised
molecular dopant is exchanged for a stable, closed-shell ion, which becomes the
stabilising counterion for the mobile charge carriers on the polymer
chains.\cite{Yamashita2019, Jacobs2021} This is depicted in
Figure~\ref{fig:first:iex}. IEx doping also provides a powerful method of
systematically studying the influence of ion size and shape on polymer charge
transport. This has enabled the identification of key factors that determine and
limit achievable electrical conductivities.\cite{Jacobs2022}

Hall effect measurements are normally a mainstay of semiconductor
characterisation, with such measurements being used to determine charge-carrier
densities and mobilities.\cite{Frederikse1967,Kamins1971,Rode1995} However, for
conjugated polymers, these measurements are not straightforward to interpret,
due to contibutions from disordered
transport.\cite{Takeya2005,Podzorov2005,Yi2016} Attempting to use the standard
single-carrier formulation, where the Hall coefficient is given by $R_H =
[qn]^{-1}$, often leads to unreliable and unphysical estimates of carrier
concentration. For instance, in their 2019 paper on IEx doping, Yamashita et al
reported charge-carrier densities of $1.4\,\times\,10^{21}$\,cm$^{-3}$ in their
devices. However, XPS and NMR measurements published by Jacobs et al in 2022
have shown that in similarly doped devices of the same material, charge-carrier
densities are actually closer to $8 \times 10^{20}$\,cm$^{-3}$ --- about half
that reported by the Hall measurements.\cite{Jacobs2022} Similarly, carrier
mobility extracted this way --- often referred to as the ``Hall mobility'' ---
is consequently an underestimate. The strong temperature dependence of the Hall
coefficient, and thus the measured carrier density, is inconsistent with
spectroscopic measurements, underscoring the unreliability of these values.

It was shown by Yi et al in 2016\cite{Yi2016} that the origin of such
discrepancies, at least in molecular crystals, can be explained by the presence
of charge carriers in localised electronic states. They suggested that localised
carriers (in contrast to carriers in extended band-like states) do not partake
in generating the transverse Hall voltage, but are still driven by its
corresponding electric field. Therefore, they partially screen the Hall voltage
leading to an underdeveloped Hall effect. Mathematically, they expressed this in
terms of two parameters: $\gamma$ and $\beta$, the fraction of band-like
carriers and ratio of localised to band-like mobility, respectively. The
subsequent reduction in the measured Hall coefficient leads to an overestimation
of carrier concentration and underestimation of carrier mobility. By taking
certain temperature limits, Yi et al were able to formulate approximate
expressions for their crystalline materials, allowing them to reliably extract
carrier density and mobility values from their measurements.

A reliable means of estimating carrier density and carrier mobility is important
for better understanding of the underpinning, fundamental charge-transport
physics. Therefore, in an effort to achieve the same for IEx-doped polymer
devices, we propose here to perform a similar analysis, thus requiring
high-resolution, temperature-dependent measurements to be undertaken. This
necessitated the construction of a new experimental system, based on the design
first published by Chen et al in 2016\cite{Chen2016} with some incremental
improvements. These improvements were both to the hardware of the system, but
also to the method of data processing used on its output, and are detailed later
on.

For the analysis itself, we first show that an expression equivalent to that
obtained from the   $\beta$-$\gamma$ analysis of the two-carrier model can also
be derived within a framework that is more appropriate for polymers --- albeit
with subtly different definitions of $\beta$ and $\gamma$. By making some
reasonable assumptions about the arguments of the expression ($\beta$ and
$\gamma$) we arrive at an approximate expression for the overall temperature
dependence of the Hall coefficient. The analysis allows us to extract reliable
values of charge-carrier density. We applied this method to highly conducting,
IEx-doped films of poly(2,5)-bis(3-alkylthiophen-2-yl)thieno[3,2-b]-thiophene
(PBTTT). We chose PBTTT due to it being a widely studied conducting conjugated
polymer model system and our ability to determine carrier concentrations in
IEx-doped PBTTT independently by spectroscopic means. 

\section{Experimental System and Data Processing}

There are a number of experimental challenges to performing
temperature-dependent Hall measurements on conjugated polymers. Polymers are
(relatively) electrically resistive materials, thus limiting the current
densities that can be injected. This results in small Hall signals and
subsequently low signal-to-noise ratios.\cite{Coropceanu2007,Karpov2017} This is
normally mitigated by using superconducting electromagnets (SCEMs) to compensate
for the low current densities. However, this is often not enough to fully
compensate for the loss of signal. Therefore, the magnetic field would normally
be ramped up and down slowly to the maximum accessible magnetic field, allowing
for sufficiently large Hall voltage values to be measured as a function of time,
and thus field. Such measurements normally take a prohibitively large amount of
time to complete and are susceptible to electromagnetic interference and slow
drift of signal. Additionally, slightly misaligned Hall electrodes on a device
can cause issues if the longitudinal voltage is at all field-dependent.
Specifically, magnetoresistance can often cause a greater change in longitudinal
voltage with respect to field strength than in the Hall voltage itself ---
particularly at low temperatures.

A solution to both of these issues is offered by AC field measurements, in which
the magnetic field is modulated (sinusoidally) on a time scale of $\sim 1$\,s.
The resulting modulation of the Hall voltage can then be detected with lock-in
techniques. This spreads noise across the entire frequency domain while
concentrating the Hall signal at a single frequency, thus significantly
increasing the signal-to-noise ratio. Furthermore, since magnetoresistance is at
least a second-order effect (i.e. $\sim O(B^2)$), no contribution from it would
be present at the field frequency.\cite{Pippard1989,Chen2016} In
Reference~\cite{Chen2016}, Chen et al implemented this with an experimental
design that used a rotating assembly of permanent magnets. These were aligned
across two plates in an alternating fashion to generate an alternating magnetic
field as it rotated. The RMS amplitude of the subsequently alternating Hall
voltage was then measured by use of a lock-in amplifier, with reference signal
provided by the analogue voltage output of a magnetometer.

In this work we made some iterative improvements upon this design (depicted in
Figure~\ref{fig:first:setup}) to allow for temperature control and to further
reduce noise. This improved system allowed us to take temperature-dependent Hall
measurements at relatively high speeds ($\sim$\,1 day for a full temperature
sweep), without magnetoresistance posing any issues, and with a greater
signal-to-noise ratio. Specifically, the following additions were made:
\begin{enumerate}

    \item A vacuum cryostat for temperature control and to prevent degradation
    of air-unstable devices;
    
    \item Minimisation of enclosed area due to loop created by Hall voltage leads, including by use of twisted-pair, cryogenic-loom wiring;
    
    \item An increased number of magnets per plate (4 instead of 2), resulting
    in a more accurately sinusoidal field and preventing the desired signal from
    being split across multiple harmonics;
    
    \item A bespoke sample-loading mechanism, allowing for devices to be swapped
    in and out with relative ease (shown in Figure~\ref{fig:first:device});
    
    \item A high-impedance ($> 1$\,T$\Omega$) voltage pre-amplifier, allowing
    for measurement of low conductivity devices (down to $< 1 $\,S/cm);
    
    \item Software-based digital PID control to stabilise the rotational
    frequency of the magnet assembly, and hence field frequency.

\end{enumerate}
Furthermore, improvements were made to the data-processing method used to
extract the Hall coefficient from raw data. This could better separate out the
Hall voltage from any Faraday-induced voltages and also allow for the sign of
the Hall coefficient to be reliably determined.

To understand this, we must consider the main voltages that contribute to the
measured output of such an AC Hall system: the Hall voltage itself, $V_H$, and
voltages induced by the Faraday effect, $V_F$, due to changing magnetic
fluxes.\cite{Chen2016} Both oscillate at the same frequency as the
field.\cite{Hall1879,Maxwell1865} To distinguish between the two contributions,
it is important to note that only $V_H$ should depend on the source-drain
current, $I$, and that $V_F$ will be $\pi / 2$ out of phase with respect to
$V_H$.\cite{Chen2016} Since a dual-phase lock-in amplifier was used, both the
in-phase ($V_x$) and $\pi / 2$ out-of-phase ($V_y$) components of the signal are
measured separately. One might then think that it is possible to directly
measure $V_H$ separately from $V_F$ using this feature. However, since the
reference signal used by the lock-in amplifier will have an arbitrary phase
offset, $\phi$, from the field experienced by the device under test (DUT), this
becomes non-trivial, with the vector voltage output becoming:
\begin{equation}
    \begin{aligned}
        \mathbf{V} = \begin{bmatrix}
            V_x\\
            V_y
        \end{bmatrix} & = \begin{bmatrix}
            \cos{\phi} & -\sin{\phi}\\
            \sin{\phi} & \cos{\phi}
        \end{bmatrix}
        \begin{bmatrix}
            V_H\\
            V_F
        \end{bmatrix}\\
        & = \begin{bmatrix}
            V_H\cos{\phi} - V_F\sin{\phi}\\
            V_H\sin{\phi} + V_F\cos{\phi}
        \end{bmatrix}.
    \end{aligned}
    \label{eqn:vectorVoltage}
\end{equation}

\begin{figure*}[htb]
    
    \centering
    \begin{subfigure}{0.32\linewidth}
        \caption{}
        \includegraphics{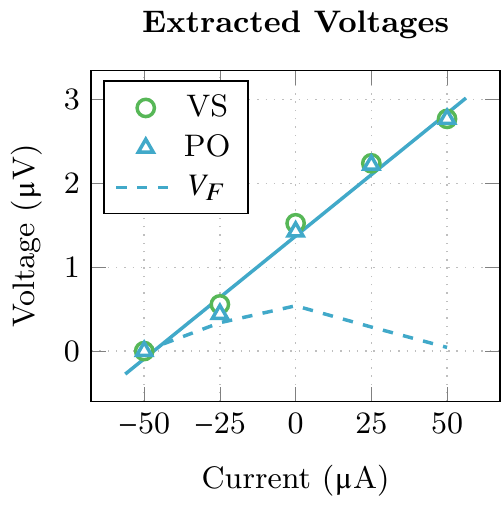}
        \label{fig:sign:voltages}
    \end{subfigure}
    \hfill
    \begin{subfigure}{0.32\linewidth}
        \caption{}
        \includegraphics{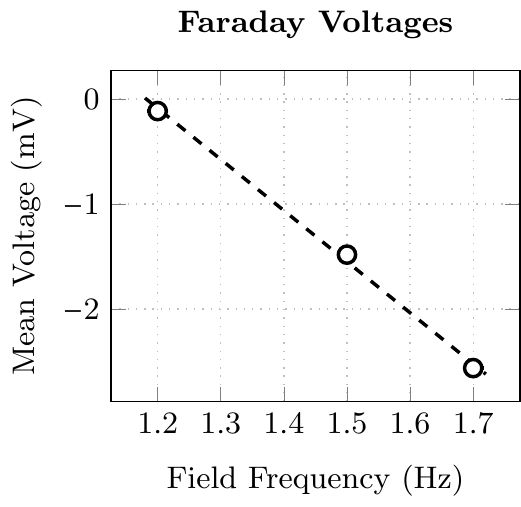}
        \label{fig:sign:faraday}
    \end{subfigure}
    \hfill
    \begin{subfigure}{0.32\linewidth}
        \caption{}
        \includegraphics{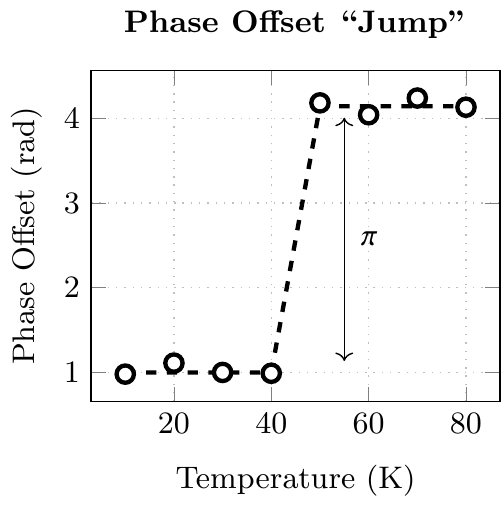}
        \label{fig:sign:jump}
    \end{subfigure}
    \caption{Data extraction from a Hall measurement:
    (\subref{fig:sign:voltages}) A seemingly trustworthy-looking measurement
    where both the vector-subtracted (VS) and phase-optimised (PO) data match
    closely and fluctuations in the Faraday voltage ($V_F$) are small relative
    to the Hall signal; (\subref{fig:sign:faraday}) Sign determination of a
    measurement, showing a negative gradient of Faraday voltage versus frequency
    indicating a positive Hall coefficient; (\subref{fig:sign:jump}) A simulated
    example of what a change in $\textnormal{sgn}(R_H)$ between two temperatures
    would look like in the calculated phase offset versus temperature --- i.e. a
    jump in phase by a value of $\pi$ between the two
    temperatures.\label{fig:sign}}

\end{figure*}

To extract $V_H$ on its own will require an assumption to be made. The
assumption made in Reference~\cite{Chen2016} was that $V_F$ is perfectly stable
for a single measurement. The result of this is that by taking the vector
differences of voltage measurements like so
\begin{equation}
    \begin{aligned}
        |\Delta \mathbf{V}| & = |\mathbf{V}(I_1) - \mathbf{V}(I_0)|\\
        & = \left|\begin{bmatrix}
            (V_H(I_1) - V_H(I_0))\cos{\phi}\\
            (V_H(I_1) - V_H(I_0))\sin{\phi}
        \end{bmatrix}\right|\\ 
        & = |\mathbf{V}_H(I_1) - \mathbf{V}_H(I_0)|\\
        & = |\Delta \mathbf{V}_H|
    \end{aligned}
    \label{eqn:vectorSubtract}
\end{equation}
one can find the magnitude of the Hall coefficient by finding the gradient of
this vector difference versus current. Which value is used as the ``zero'' value
is an arbitrary choice. For instance, one could simply take the first
measurement in a sweep as being $\mathbf{V}(I_0)$ and subtract it from all measured
points.

This method is quick and can be performed live as data is acquired. However, the
downside is that any noise in $V_F$ ends up being incorporated into the
extracted $V_H$ signal. The assumption that $V_F$ is constant is also not
entirely physical, as the moving parts in the system are liable to cause $V_F$
to fluctuate throughout a measurement via mechanical vibrations.

Alternatively,  one could assume that the phase offset between the reference
signal and DUT-experienced magnetic field is constant. If the position of the
magnetometer sensor is fixed, then this would seem to be a more physical
assumption. Further to this, if one assumes that $V_F$ fluctuates around a
constant value, then it is possible to use numerical methods in an attempt to
determine the phase offset, and separate $\mathbf{V}$ back into $V_H$ and $V_F$.

This can be formulated as a minimisation problem by defining a minimisation
parameter, $P(\phi)$, to be
\begin{equation}
    P(\phi) = \left|\dfrac{M_y(\phi)}{M_x(\phi)}\right|
    \label{eqn:minimisation}
\end{equation}
where $M_{n}(\phi)$ is the gradient resulting from a linear fit of $V_n$ vs $I$
after rotating $\mathbf{V}$ through a phase $\phi$. The value of $\phi$ that
minimises $P(\phi)$ should be the phase needed to rotate $\mathbf{V}$ through to
separate out the Hall and Faraday components, since one would expect that
$\av{\partial V_F / \partial I} \approx 0$ and $|\av{\partial V_H / \partial I}|
\gg |\av{\partial V_F / \partial I}|$, where $\av{...}$ indicates a mean value.

Both methods have drawbacks. Vector subtraction (VS) is inherently noisier while
phase optimisation (PO) may home in on the wrong value of $\phi$ if $V_F$
happens to have some sort of correlation with current (either by random chance
or otherwise). Therefore, for each measurement performed using this system, both
methods were used and their outputs compared. This provided a means of testing
the reliability of a measurement. If the two outputs differed significantly or
the extracted $V_F$ signal had large fluctuations relative to the $V_H$ signal,
then the measurement was deemed unreliable. An example of a typical
``trustworthy'' measurement is shown in Figure~\ref{fig:sign:voltages}.

In both methods explored thus far, it is only possible to tell the magnitude of
the Hall coefficient --- not its polarity. For VS, this is because only the
magnitude of a vector is being found. For PO, this is because there is no way to
tell whether $\phi$ or $\phi + \pi$ is the ``correct'' offset without more
information. However, it should be possible to extract the sign of the Hall
coefficient by checking how $V_y$ (Faraday) varies with frequency relative to
how $V_x$ (Hall) varies with current. When properly aligned, the $x$ and $y$
components should be
\begin{equation}
    \begin{aligned}
        V_x &= R_H B_z I \sin(\omega t) / d\\
        V_y &= - \omega \Omega B_z \sin(\omega t + \pi / 2)
    \end{aligned}
    \label{eqn:voltageComponents}
\end{equation}
where $R_H$ is the Hall coefficient, $\Omega$ is the cross-sectional area
enclosed by the loop created by the Hall measurement wires, $B_z$ is the
perpendicular magnetic flux density amplitude, $I$ is current, $\omega$ is the
angular frequency of the field, $t$ is time and $d$ is the thickness of the DUT.
If $R_H$ is positive, then the gradient of $V_x$ vs $I$ should have the opposite
sign to the gradient of $V_y$ vs $\omega$. If it is negative then they will have
the same sign. Therefore, after rotating through the determined value of $\phi$,
the sign of $R_H$ can be determined from:
\begin{equation}
    \textrm{sgn} (R_H) = -\textrm{sgn} \left( \dfrac{\partial V_y}{\partial \omega} \dfrac{\partial V_x}{\partial I} \right).
\end{equation}
Therefore, by performing frequency-dependent measurements one can extract the
sign of the Hall coefficient as well as its magnitude.

This method was simplified for these measurements by always taking the value of
$\phi$ that yields a positive Hall coefficient, then determining the sign by use
of $\textrm{sgn}(R_H) = -\textrm{sgn}(\partial V_y / \partial \omega)$. An
example of this is shown in Figure \ref{fig:sign:faraday}. Further to this, only
one frequency sweep at one temperature would be necessary, as any changes in
sign of the Hall coefficient should result in whether $\phi$ or $\phi + \pi$ is
the positive result changing. This would result in a ``jump'' in the selected
phase offset by a value of $\pi$ between measurements. An illustration of what
this would look like is shown in Figure~\ref{fig:sign:jump}. Therefore, if the
sign is known at one temperature, the sign at all other temperatures can be
determined by spotting these $\pi$-sized discontinuities in the phase offset
versus temperature.

Given recent observations of n-type behaviour in some p-doped polymer Hall
measurements\cite{Liang2021}, having the ability to determine the sign of $R_H$
was considered to be important for these measurements. This method was used on
the PBTTT devices measured here to determine that the Hall coefficient was
positive at all temperatures --- as was expected.

While the devices measured in this work did not exhibit any noticeable frequency
dependence in their Hall coefficients, it is conceivable that other devices may.
In such cases, since frequency-dependent measurements will be necessary anyway,
it is worth noting that extrapolating to zero frequency should effectively
remove any influence of Faraday induction on the measured Hall signal (since
$V_F \propto \omega$).

\begin{figure*}[bt]

    \centering
    \begin{subfigure}{0.32\linewidth}
        \caption{}
        \includegraphics{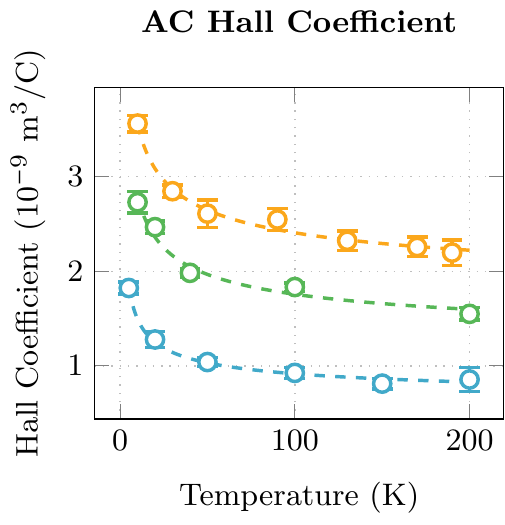}
        \label{fig:data:achall}
    \end{subfigure}
    \hfill
    \begin{subfigure}{0.32\linewidth}
        \caption{}
        \includegraphics{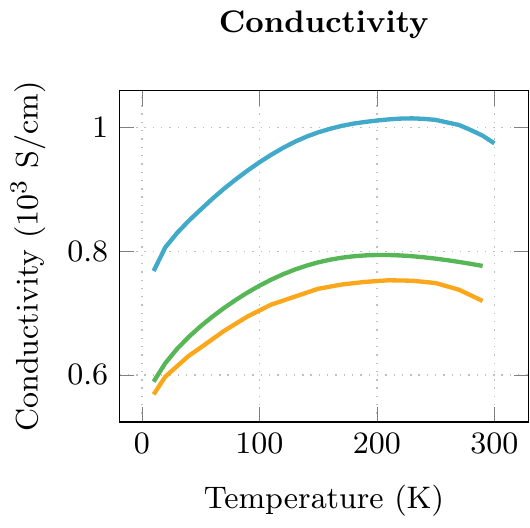}
        \label{fig:data:accond}
    \end{subfigure}
    \hfill
    \begin{subfigure}{0.32\linewidth}
        \caption{}
        \includegraphics{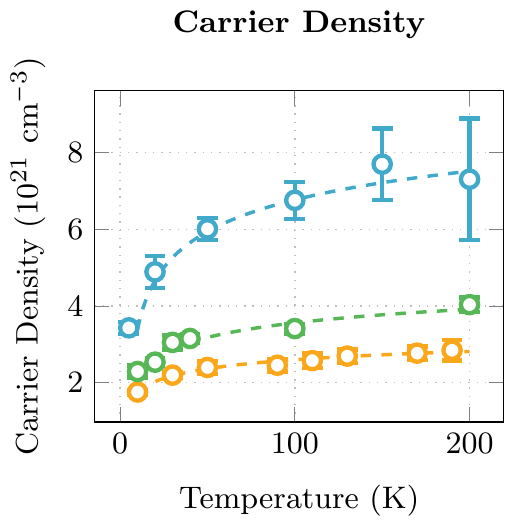}
        \label{fig:data:acdensity}
    \end{subfigure}\\
    \begin{subfigure}{0.32\linewidth}
        \caption{}
        \includegraphics{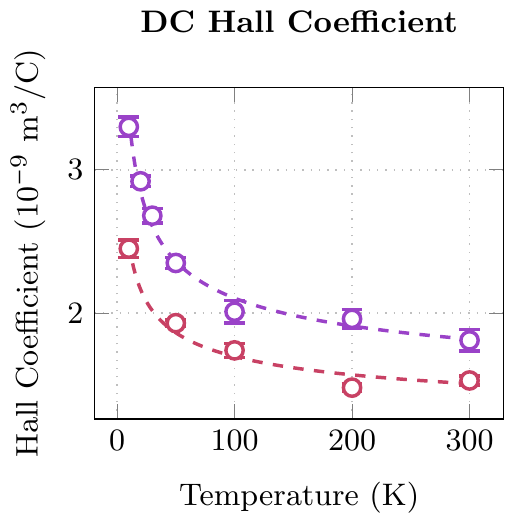}
        \label{fig:data:dchall}
    \end{subfigure}
    \hfill
    \begin{subfigure}{0.32\linewidth}
        \caption{}
        \includegraphics{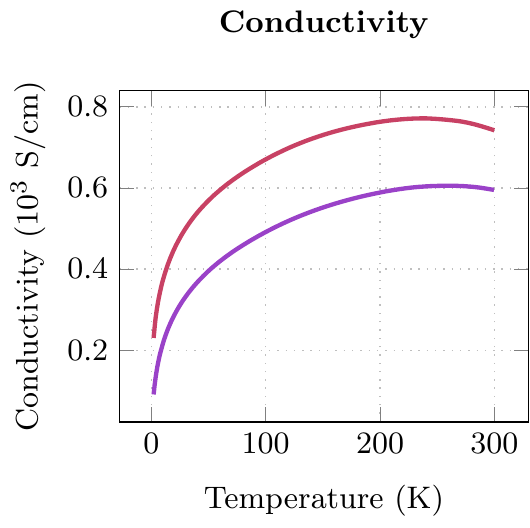}
        \label{fig:data:dccond}
    \end{subfigure}
    \hfill
    \begin{subfigure}{0.32\linewidth}
        \caption{}
        \includegraphics{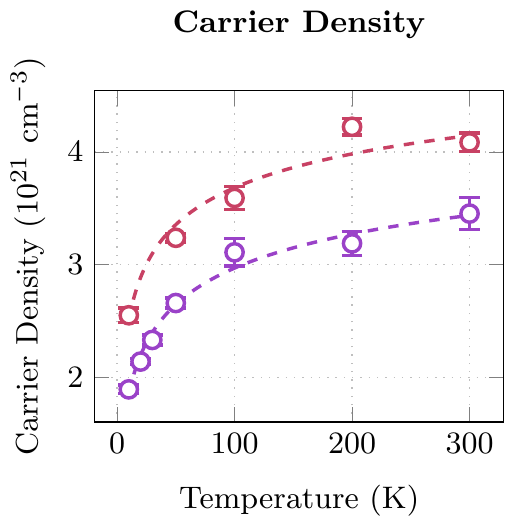}
        \label{fig:data:dcdensity}
    \end{subfigure}
    \caption{Plots of temperature-dependent data acquired from 5 IEx-doped Hall
    bar devices, 3 measured using the AC Hall system and 2 confirmatory
    measurements performed using a DC Hall system: (\subref{fig:data:achall} +
    \subref{fig:data:dchall}) Hall coefficient with fitted ``guides for the
    eye'' (dashed); (\subref{fig:data:accond} + \subref{fig:data:dccond})
    Conductivity, (\subref{fig:data:acdensity} + \subref{fig:data:dcdensity})
    Hall carrier density with fitted ``guides for the eye''
    (dashed).\label{fig:data}}

\end{figure*}

\section{PBTTT Hall Effect Measurements}

Measurements were performed on devices of PBTTT-C$_{14}$, fabricated and doped
using the methods described in Reference~\cite{Jacobs2021} using FeCl$_3$ as the
molecular dopant and BMP TFSI as the electrolyte for IEx. The polymer films
(thickness 40\,nm) were patterned into a Hall bar shape (rectangle of length
900\,$\muup$m, width 100\,$\muup$m, with two pairs of bottom-contact Hall
electrodes spaced equidistantly at 300\,$\muup$m intervals along the channel, as
shown in Figure~\ref{fig:first:device}) using the methods described in
Reference~\cite{Chang2010}. Three different films, doped to different levels of
conductivity, were measured in the AC Hall system. Their room temperature
conductivities (987\,S/cm, 776\,S/cm, and 719\,S/cm) were high, with their
temperature dependencies revealing metallic transport signatures. Specifically,
at room temperatures, their conductivities decrease with increasing temperature
and, at temperatures as low as 10\,K, their conductivities retain values that
are between $70-80\%$ of those at room temperature. As control measurements, two
further devices were measured in a standard DC Hall system with a 12\,T SCEM.
The results of the AC Hall measurements are shown in
Figures~\ref{fig:data:achall}~to~\ref{fig:data:acdensity} and DC in
Figures~\ref{fig:data:dchall}~to~\ref{fig:data:dcdensity}. The undoped thickness
of the films was used to calculate $R_H$, as it should allow us to more readily
compare results to other known quantities pertinent to the undoped film (such as
monomer and polymer chain density).

For the AC measurements, temperature was swept from 10\,K to 300\,K in steps of
10\,K. Previous tests had shown no hysteresis with temperature, thus it was only
necessary to sweep in one direction. At each temperature step, the system would
be left to stabilise for at least 30 minutes before performing any measurements.
After stabilisation at each temperature, conductivity was measured, whereas Hall
measurements were only performed (after conductivity) at a subset of steps
(typically 10\,K, 20\,K, 30\,K, 40\,K, 50\,K, 100\,K, 150\,K, and 200\,K). This
was because the low-temperature measurements were thought to be the most
important for later analysis with values $> 50$\,K being less important.

The conductivity measurements were four-wire measurements performed using a
Keithley 2450 source-measure unit (SMU), using its independent force and sense
probes. For the Hall measurements, the voltage signal was extracted by use of a
Stanford Research Systems SR830 lock-in amplifier via a SR551 voltage
pre-amplifier ($> 1$\,T$\Omega$ input impedance) using a 30\,s time constant
(integration time) and $\pm 500$\,$\muup$V sensitivity range, with current being
injected by the previously mentioned SMU (now operating in two-wire mode).

The choice of frequency values to use was also important. Lower frequencies
would require the use of longer integration times, as well as having to compete
with greater $1/f$ noise. However, higher frequencies would result in greater
noise from Faraday-induced voltages. It was found that a frequency of 1.2\,Hz
achieved a good balance between these competing factors, while still being
reliably achievable using the DC motor (i.e. the motor and PID control were
found to reliably and stably achieve this speed without stalling). Furthermore,
minimising the area enclosed by the Hall voltage measurement loop --- for
instance by ensuring wire bonds were not overly long --- helped minimise Faraday
induction. For the room-temperature, frequency-dependent measurements,
frequencies of 1.2\,Hz, 1.5\,Hz, and 1.7\,Hz were used. Similarly, these values
were chosen due to them being reliably achievable.

The data generally reveal good consistency between AC and DC measurements; both
setups measure the same temperature dependence, thus showing the AC measurements
in our new experimental setup to be reliable. The measurements show a pronounced
negative correlation of $R_H$ with temperature, with this dependence becoming
stronger at lower temperatures. This is consistent with what one should expect:
as temperature decreases, the localised carriers (which move by a
thermally-activated hopping transport mechanism) should become less mobile and
thus less able to screen out the Hall voltage. The result of this screening is
that at high temperatures the charge-carrier densities extracted by assuming the
simple $R_H = [qn]^{-1}$ relation (plotted in
Figures~\ref{fig:data:acdensity}~and~\ref{fig:data:dcdensity}) are clearly
spurious. For instance, at 200\,K the largest value of $\sim$\,$8 \times
10^{21}$\,cm$^{-3}$ represents a density of $\sim$\,$8$ carriers per monomer,
since PBTTT typically has monomer densities $\sim$\,$1 \times
10^{21}$\,cm$^{-3}$. That is, each monomer unit would need to be in a stable
$+8$ oxidation state, which is chemically impossible.

\section{Model Formulation}

To develop a reliable method for analysing such Hall data, we first discuss the
model for the under-developed Hall effect proposed in Reference~\cite{Yi2016}.
This model, devised for crystalline organic small-molecule semiconductors,
assumes the existence of two uniform populations of carriers: ``band-like''
carriers and localised, ``hopping'' carriers, denoted here with subscripts $b$
and $h$ respectively. By assuming that only band-like carriers experience a
non-negligible contribution from the $\mathbf{v} \times \mathbf{B}$ term in the
Lorentz force (for instance, one could argue that they have a negligibly small
drift velocity), the Hall coefficient can be expressed as
\begin{equation}
    R_H = \dfrac{n_b \mu_b^2}{q(n_b \mu_b + n_h \mu_h)^2}
    \label{eqn:RHYi2016}
\end{equation}
where $q$ is the charge on each carrier, $n_{b,h}$ are the respective
charge-carrier densities, and $\mu_{b,h}$ are the respective charge-carrier
mobilities. Then by defining the two parameters, $\beta$ and $\gamma$, with
definitions
\begin{equation}
    \beta = \dfrac{\mu_h}{\mu_b} \qquad \gamma = \dfrac{n_b}{n_b + n_h}
    \label{eqn:betaGammaDefinitions}
\end{equation}
the Hall coefficient can be written as
\begin{equation}
    R_H = \dfrac{1}{q n_b} \left[ \dfrac{\gamma}{\gamma + (1 - \gamma)\beta} \right]^2.
    \label{eqn:betaGammaHall}
\end{equation}

We now show that a very similar expression can be derived from a more
generalised approach, which may be more suited to polymers. In this approach,
instead of two uniform populations, we think of carriers as existing in a single
population distributed on a continuous scale from completely localised to
completely band-like. In an attempt to account for this, we assign each carrier
a continuous and dimensionless ``magnetic coupling parameter'', $g$, where $0
\leq g \leq 1$, to quantify how strongly each carrier ``feels'' the magnetic
term in the Lorentz force:
\begin{equation}
    \mathbf{F} = q \left[ \mathbf{E} + g \left( \mathbf{v} \times \mathbf{B} \right) \right].
\end{equation}

By performing the usual derivation for the Hall effect (i.e. balancing
transverse currents to zero)\cite{Kittel2018} and integrating over the total
population of carriers, the Hall coefficient becomes
\begin{equation}
    R_H = \dfrac{1}{q n_t} \dfrac{\int_0^1 g \mu_{_B}^2(g) f(g)\,\textnormal{d}g}{\left[ \int_0^1 \mu_{_E}(g) f(g)\,\textnormal{d}g \right]^2}
    \label{eqn:generalised}
\end{equation}
where $n_t$ is the total charge-carrier density, and $f(g)$ is the distribution
function describing the proportion of carriers that have a given value of $g$.
It is worth noting that in this expression, we are implicitly assuming that
carriers with different values of $g$ contribute to the conduction as channels
in parallel. The subscripts of $\mu$ --- $E$ and $B$ --- are introduced to keep
track of which mobility values pertain to acceleration from the electric and
magnetic parts of the Lorentz force respectively. For instance, if one were to
assume that electric fields are able to accelerate carriers for their entire
time of flight, but only for a portion of it for magnetic fields, then a
carriers average mobility as ``seen'' by magnetic fields will not necessarily be
the same as ``seen'' by electric fields (as they will be averaged over different
periods of time).

Regardless of assumption, it can be seen that the integrals in
Equation~\ref{eqn:generalised} are over the entire domain of, and contain, the
distribution function, $f(g)$. Therefore, they can be written as expectation
values, allowing us to write
\begin{equation}
    R_H = \dfrac{1}{q n_t} \dfrac{\av{g \mu^2_{_B}}}{\av{\mu_{_E}}^2}
    \label{eqn:expectation}
\end{equation}
thus showing that the ``normal'' Hall coefficient ($[q n_t]^{-1}$) is reduced by
an interplay between the most-delocalised carriers providing the greatest
contribution to the numerator, $\av{g \mu^2_{_B}}$, due to being weighted by
$g$, and the least-delocalised carriers contributing significantly only to the
denominator, $\av{\mu_{_E}}^2$.

Equations~\ref{eqn:generalised}~and~\ref{eqn:expectation} should be thought of
as general (albeit abstract) expressions for describing the under-developed Hall
effect in single-carrier, partly disordered materials. The intricacies of how it
applies to a specific material, then, are to be found in what functional form
its important parameters (i.e. $\mu_{_{E,B}}(g)$ and $f(g)$) are assumed to
take. For instance, if one were to assume that
\begin{equation}
    \begin{aligned}
    n_t f(g) &= n_b \delta(g - 1) + n_h \delta(g)\\
    \mu_{_E} &= \mu_{_B} = \left\{ \begin{matrix} \mu_h & \textnormal{for}\,g = 0\\ \mu_b & \textnormal{for}\,g = 1 \end{matrix} \right.
    \end{aligned}
\end{equation}
where $\delta(x)$ is the Dirac delta distribution, then the expression will
reduce back to that in Equation~\ref{eqn:betaGammaHall} with $\gamma = f(1) /
(f(0) + f(1))$ and $\beta = \mu(0) / \mu(1)$. Therefore, the two-carrier model
can be thought of as being the result of binning all carriers into having either
$g = 0$ or $g = 1$ (i.e. being either fully localised or fully delocalised).

However, in polymers such as PBTTT, charge transport is often thought of as
occuring along conductive pathways, or ``fibrils'', comprising different
sections where carriers undergo different modes of transport (typically
band-like, hopping, and disordered-metallic).\cite{Kaiser1990} Properties, such
as resistivity, are then predicted by adding these contributions together in
series, weighted by the fraction of the fibril they account for. Therefore, for
the Hall coefficient, we took a similar approach by assuming that $g$ comes
about due to a carrier only spending a proportion $g$ of its time of flight
moving via mode(s) of transport under which the magnetic field is able to
deflect it.

To model this, we define $\av{\mu_d}$ to be the average mobility of carriers
when they are undergoing such ``deflectable'' transport, and $\av{\mu_n}$ to be
that when they are undergoing ``non-deflectable'' transport. Therefore, by
assuming that carriers are always being driven be electric fields, regardless of
transport mechanism, we can write $\mu_{_E}$ as the weighted average of both
average mobilities:
\begin{equation}
    \mu_{_E}(g) = g \av{\mu_d} + (1 - g) \av{\mu_n}.
\end{equation}
Similarly, by assuming that magnetic fields effectively only act on the carrier
when it is travelling via deflectable mechanisms, we equate $\mu_{_B}$ to the
average deflectable mobility only:
\begin{equation}
    \mu_{_B}(g) = \av{\mu_d}.
\end{equation}
Then, by defining $\int_0^1 g f(g) \textnormal{d}g = \av{g}$, and $b =
\av{\mu_n} / \av{\mu_d}$, the Hall coefficient can now be expressed as
\begin{equation}
    R_H = \dfrac{1}{q \av{n_d}} \left[ \dfrac{\av{g}}{\av{g} + (1 - \av{g})b} \right]^2
    \label{eqn:hallNoFunction}
\end{equation}
where $\av{n_d} = \av{g} n_t$ is the average effective deflectable carrier
density --- that is the average density of carriers that are deflectable at any
given moment in time. This expression is very similar to that derivable from
Reference~\cite{Yi2016}, with the only difference being that $\gamma$ and
$\beta$ have been substituted with $\av{g}$ and $b$ respectively.

This implies that by allowing carriers to exist anywhere in a linear
interpolation between the fully-localised and fully-delocalised states
previously assumed, one still arrives at the same functional form --- albeit
with subtly different definitions to its arguments. That is, the value of
$\av{g}$ or $\gamma$ may be thought of as an average measure of the overall
delocalisation of carriers, as opposed to the fraction of carriers that are
fully delocalised. Therefore, the two-carrier functional form from
Reference~\cite{Yi2016} should be usable for polymers so long as these different
definitions are kept in mind. To emphasise this, from this point onward, we
shall use the notation $\av{g}$ and $b$ instead of $\gamma$ and $\beta$.

\section{Relating to Physical Quantities}

\begin{figure*}[htb]
    
    \centering
    \begin{subfigure}[t]{0.32\linewidth}
        \caption{}
        \includegraphics{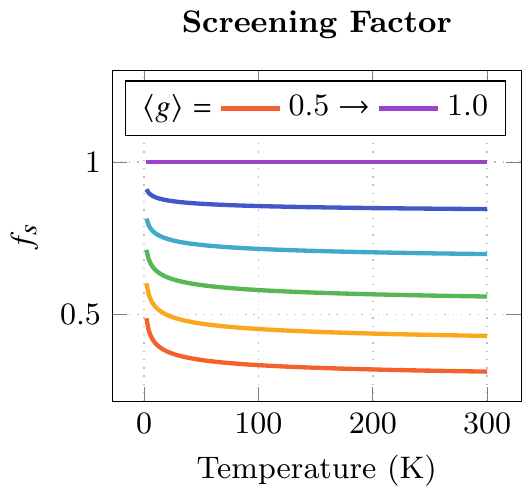}
        \label{fig:screenVTemp:screen}
    \end{subfigure}
    \hfill
    \begin{subfigure}[t]{0.32\linewidth}
        \caption{}
        \includegraphics{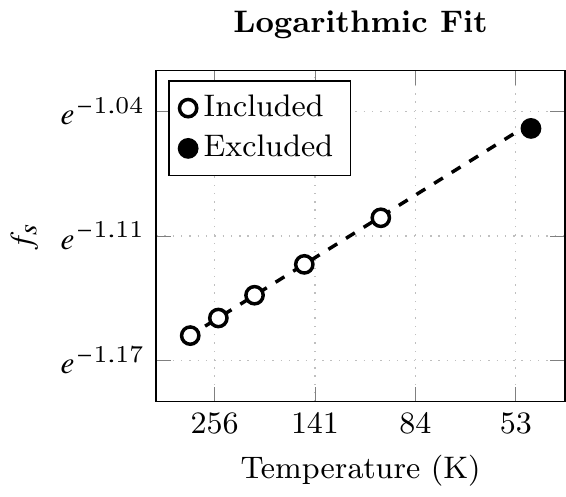}
        \label{fig:screenVTemp:logfit}
    \end{subfigure}
    \hfill
    \begin{subfigure}[t]{0.32\linewidth}
        \caption{}
        \includegraphics{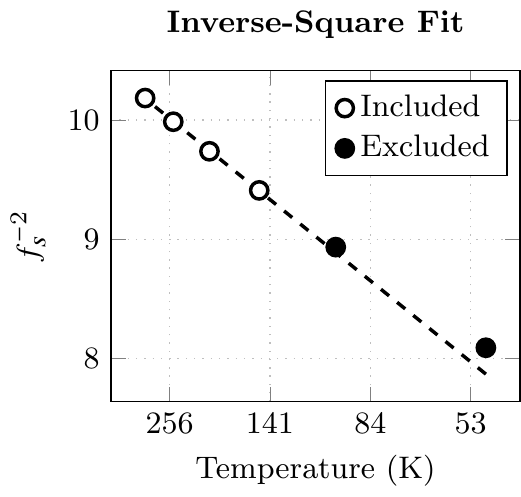}
        \label{fig:screenVTemp:invfit}
    \end{subfigure}
    \caption{Various plots of the theoretical temperature dependence of the Hall
    screening factor with $D = 0.25$, $T_* = 1$\,K, $b_{\infty} = 1$, and
    $\av{g} = 0.5$ unless stated otherwise: (\subref{fig:screenVTemp:screen})
    Plotted as a function of temperature for $0.5 \leq \av{g} \leq 1.0$,
    (\subref{fig:screenVTemp:logfit}) Plotted logarithmically against $T^{-D}$
    demonstrating a linear relationship for small $T^{-D}$ or large $T$,
    (\subref{fig:screenVTemp:invfit}) Plotted as its inverse square against
    $T^{-D}$ again showing a linear relationship for small $T^{-D}$ or large
    $T$. \label{fig:screenVTemp}}

\end{figure*}

With just two independent experimental measurements at each temperature ($R_H$
and $\sigma$) it is not possible to extract the three model parameters
($\av{n_d}$, $\av{g}$ and $b$) unambiguously. Therefore, it is necessary to make
further assumptions. Since the polymer systems in question are highly doped --
implying that the Fermi energy is large compared to $k_B T$ --- it is unlikely
that a change in temperature will affect the distribution of carriers across
different modes of transport much. Therefore, we assume $\av{g}$ to be mainly a
function of carrier concentration, and independent of temperature. On the other
hand $b$, being a ratio of mobility values, is likely to be a strong function of
temperature only:
\begin{equation}
    \av{g} \approx \av{g} (n_t) \qquad b \approx b(T).
\end{equation}

Finding an appropriate functional form for $b(T)$ requires us to consider which
scenarios, that a charge carrier might find itself in, would result in
deflectable and non-deflectable transport. The simplest pair of contrasting
scenarios would be when a carrier is undergoing localised,
variable-range-hopping-like transport versus delocalised, band-like transport.
In the former case, such carriers would not have a coherent enough wavevector to
be deflected by a magnetic field, whereas in the latter case they would. Thus,
it may be tempting to suggest that $\av{\mu_n}$ should arise from hopping-like
transport and $\av{\mu_d}$ from band-like transport. However, there are other
factors in polymers that may cause a carrier to become non-deflectable.

For instance, polarons in different regions of the polymer with differing
amounts of structural disorder, will have differing degrees of delocalisation,
while quite possibly still having band-like mobilities. Similarly, the effect of
transient (de)localisation may be such that otherwise localised carriers, being
excited into band-like states temporarily, do not remain in such delocalised
states long enough to be fully deflected by magnetic fields. Whatever the case,
it is clear that which mobility contributions should be categorised into the
``deflectable'' and ``non-deflectable'' categories is not as obvious as just
deflectable being band-like and non-deflectable being hopping-like. Therefore, a
better approach may be a semi-empirical one, where we choose a functional form
of $b(T)$ based on the behaviour of the measured data.

Since the functional form, when incorporated into
Equation~\ref{eqn:hallNoFunction}, will take the form $\sim \left[ 1 + b(T)
\right]^{-2}$ we will want a function for $b(T)$ that is small with a strong
temperature dependence at low temperatures, while plateauing at higher
temperatures. A good fit for this would be an exponential function, similar to
that used for variable-range hopping:
\begin{equation}
    b(T) \approx b_\infty \exp{\left\{ - \left[ \dfrac{T_*}{T} \right]^D \right\}}
\end{equation}
where $b_\infty$, $T_*$, and $D$ are all positive constants to be fitted. At low
$T$ ($T \lesssim T_*$) this function has a small value ($b(T) \rightarrow 0$ as
$T \rightarrow 0$) with a strong temperature dependence, and at high
temperatures ($T \gg T_*$) it plateaus asymptotically towards $b_\infty$.

Taking this functional form for $b(T)$ yields an approximate expression for the
Hall coefficient given by:
\begin{equation}
    R_H(T) \approx R_H^0 \left[ \dfrac{ \av{g} }{ \av{g} + (1 - \av{g}) b_\infty e^{-\left(T_* / T\right)^D}} \right]^{2}
    \label{eqn:fullHall}
\end{equation}
where $R_H^0$ has been introduced as the ``deflectable Hall coefficient'' given
by $R_H^0 = [q \av{n_d}]^{-1}$. For convenience, we also define $f_s(T) = R_H(T)
/ R_H^0$ as the ``screening factor''. To illustrate the expected dependence of
the Hall coefficient on temperature, we choose an example with parameters
$b_\infty = 1$, $T_* = 1$\,K, $D = 0.25$, and $0.5 \leq \av{g} \leq 1.0$,
allowing us to plot $f_s$ against temperature in
Figure~\ref{fig:screenVTemp:screen}. Qualitatively, its behaviour is in good
agreement with the measured data, providing assurance that it might be possible
to fit the data to this model. 

It should now be possible to attempt extracting the values of the functions
parameters by fitting to temperature-dependent Hall coefficient data. However,
it is still important to note that in this model, there are 5 parameters:

\begin{tabular}{rll} 
    1. & $T_*$      & The mobility-ratio temperature coefficient;\\[5pt]
    2. & $D$        & The mobility-ratio temperature exponent;\\[5pt]
    3. & $b_\infty$ & The mobility ratio value as $T \rightarrow \infty$;\\[5pt]
    4. & $\av{g}$   & The mean magnetic coupling parameter;\\[5pt]
    5. & $R_H^0$    & The deflectable Hall coefficient.\\[5pt]
\end{tabular}\\
Therefore, to achieve a reliable fit it was important to constrain some of their
values via alternative methods. This was done by approximating the full
expression in certain temperature limits. For instance, when $b(T) \gg \av{g} /
(1 - \av{g})$ (i.e. when non-deflectable transport is dominating), the
expression can be approximated as
\begin{equation}
    R_H \approx R_H^0 \left[ \dfrac{\av{g}}{b_\infty (1 - \av{g})} \right]^2 \exp{\left\{ 2 \left[ \dfrac{T_*}{T} \right]^D \right\}}
    \label{eqn:approxHall}
\end{equation}
which was used to fit the ``guides for the eye'' in
Figures~\ref{fig:data:achall}~and~\ref{fig:data:dchall}. Following this, $D$ can
now be estimated by finding the value that gives the most linear fit of
$\ln{(R_H)}$ against $T^{-D}$, allowing $T_*$ to then be estimated by use of
\begin{equation}
    T_* \approx \left[ \dfrac{1}{2} \dfrac{\partial \ln{(R_H)}}{\partial T^{-D}} \right]^{\frac{1}{D}}
\end{equation}
which can be calculated using a linear fit, an example of which is shown in
Figure~\ref{fig:screenVTemp:logfit}.

Following from this, if one assumes to be in the limit of $T \gg T_*$, then
$R_H^{0}$ can be estimated by fitting $R_H^{-1/2}$ against $T^{-D}$ to yield a
linear relation of
\begin{equation}
    \begin{aligned}
        R_H^{-1/2} & \approx MT^{-D} + C\\
        M & = - \left(R_H^0\right)^{-\frac{1}{2}} \left( \av{g}^{-1} - 1 \right) b_\infty T_*^D\\
        C & = \left(R_H^0\right)^{-\frac{1}{2}} - \dfrac{M}{T_*^D}
    \end{aligned}
\end{equation}
thus allowing for $R_H^{0}$ to be estimated from
\begin{equation}
    R_H^{0} \approx \left[ C + \dfrac{M}{T_*^D} \right]^{-2}
\end{equation}
an example fit of which is shown in Figure~\ref{fig:screenVTemp:invfit}.

At this point, three out of five parameters have been estimated, thus reducing
the fitting problem to a relatively simple two-variable one. However, $\av{g}$
and $b_\infty$ are not uniquely solvable. This can be seen by dividing
Equation~\ref{eqn:fullHall} through by $\av{g}^2$ to give:
\begin{equation}
    R_H(T) \approx R_H^0 \left[ 1 + h e^{-(T_*/T)^D} \right]^{-2}
\end{equation}
where we have defined $h$ to be
\begin{equation}
    h = \dfrac{1 - \av{g}}{\av{g}} b_\infty.
\end{equation}
Different value pairs of $\av{g}$ and $b_\infty$ can be chosen and, so long as
they produce the same value of $h$, they will produce the same fit. Therefore,
the true fitting parameter is $h$. By considering the meanings of the parameters
that make up $h$, one can arrive at
\begin{equation}
    h e^{-(T_*/T)^D} = \dfrac{\sigma_n(T)}{\sigma_d(T)}
    \label{eqn:condRatio}
\end{equation}
where $\sigma_{n,d}$ are the temperature-dependent contributions to conductivity
from non-deflectable and deflectable transport respectively. We can then
rearrange to give the fraction of conduction that is deflectable, as a function
of temperature, the result of which can be seen to equal the square root of the
screening factor, $f_s$:
\begin{equation}
    d(T) = \dfrac{\sigma_d}{\sigma_d + \sigma_n} = \left[ 1 + h e^{-(T_*/T)^D} \right]^{-1} = \sqrt{f_s}.
    \label{eqn:defFrac}
\end{equation}
Furthermore, Equation~\ref{eqn:condRatio} allows us to calculate $\av{\mu_d}$ as
a function of temperature by use of
\begin{equation}
    \av{\mu_d}(T) = \dfrac{\sigma(T) R_H^0}{1 + h e^{-(T_*/T)^D}}
    \label{eqn:defMob}
\end{equation}
where $\sigma(T)$ is the measured temperature-dependent conductivity.

By considering temperature limits, it can be seen that
Equations~\ref{eqn:condRatio} taken in the limit of $T \gg T_*$ --- that is when
it has plateaued --- simply gives $h$. Therefore, $h$ on its own can be a useful
parameter, since Equation~\ref{eqn:defFrac} can then be estimated by: 
\begin{equation}
    d(T \gg T_*) \approx (1 + h)^{-1}.
\end{equation}
\setlength{\extrarowheight}{4pt}{%
\begin{figure*}[htb]
    \centering
    \centering
    \begin{subfigure}[c]{0.59\linewidth}
        \includegraphics{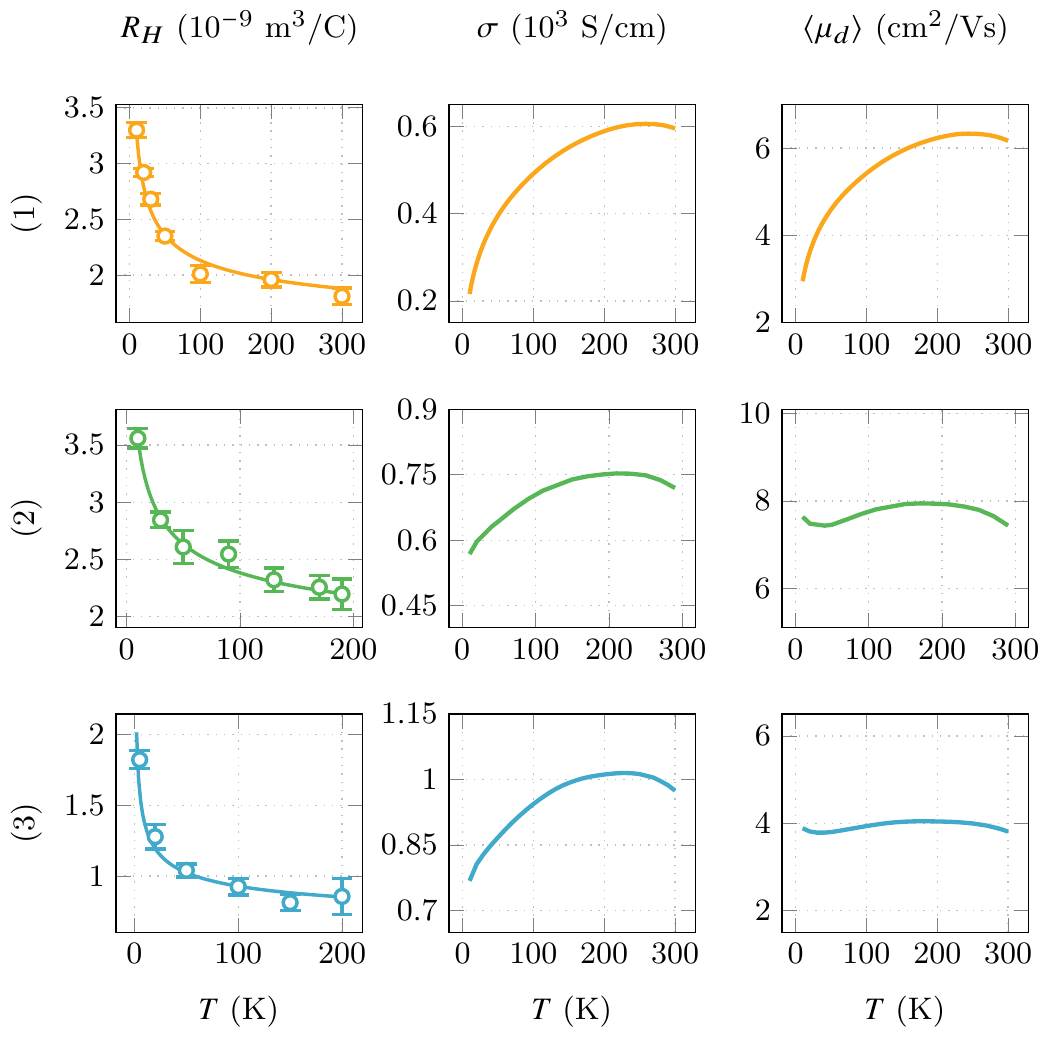}
    \end{subfigure}
    \hfill
    \begin{subfigure}[c]{0.39\linewidth}
        \begin{ruledtabular}
            \begin{tabular}{crr}
                                           & $D$                              & $T_*$ (K)                        \\[\extrarowheight] \hline
                1 \crule[plot2]{15pt}{6pt} & $0.299 \pm 0.019$                & $1.34 \pm 0.10$                  \\[\extrarowheight]   
                2 \crule[plot3]{15pt}{6pt} & $0.239 \pm 0.018$                & $1.35 \pm 0.20$                  \\[\extrarowheight]  
                3 \crule[plot4]{15pt}{6pt} & $0.226 \pm 0.030$                & $1.31 \pm 0.21$                  \\[\extrarowheight] \hline
                                           & $R_H^{0}$ (m$^3$/C)              & $\av{n_d}$ (cm$^{-3}$)           \\[\extrarowheight] \hline
                1 \crule[plot2]{15pt}{6pt} & $(5.72 \pm 0.29) \times 10^{-8}$ & $(1.09 \pm 0.05) \times 10^{20}$ \\[\extrarowheight]
                2 \crule[plot3]{15pt}{6pt} & $(5.08 \pm 0.48) \times 10^{-8}$ & $(1.23 \pm 0.12) \times 10^{20}$ \\[\extrarowheight]
                3 \crule[plot4]{15pt}{6pt} & $(1.87 \pm 0.10) \times 10^{-8}$ & $(3.34 \pm 0.18) \times 10^{20}$ \\[\extrarowheight] \hline
                                           & $1/[1 + h]$ (\%)                 & $\av{\mu_d}$ (cm$^2$/Vs)         \\[\extrarowheight] \hline
                1 \crule[plot2]{15pt}{6pt} & $15.4 \pm 0.21$                  & $2.95 - 6.33$                    \\[\extrarowheight]
                2 \crule[plot3]{15pt}{6pt} & $16.2 \pm 0.37$                  & $7.43 - 7.94$                    \\[\extrarowheight]
                3 \crule[plot4]{15pt}{6pt} & $16.4 \pm 0.92$                  & $3.78 - 4.05$                    \\[\extrarowheight]
            \end{tabular}
        \end{ruledtabular}
    \end{subfigure}
    \caption{The results of fitting the Hall data acquired from three IEx-doped
    PBTTT hall bars, with plots of Hall coefficient ($R_H$), conductivity
    ($\sigma$) and mean deflectable mobility ($\av{\mu_d}$) on the left, and a
    table of fitted parameters on the right.}
    \label{fig:mainTable}

\end{figure*}}
If $T_*$ is much smaller than room temperature, then this value should
accurately estimate the fraction of conductivity that is deflectable at room
temperature.

Estimating $h$ can be done with the approximate expression in
Equation~\ref{eqn:approxHall} by realising that the prefactor to its exponential
is equal to $R_H^0 h^{-2}$. Therefore, we have managed to estimate all fitting
parameters, making a fit of the full expression relatively simple. However, the
limits in which these approximations apply cannot be determined until the values
they approximate have been calculated. Therefore, the estimates they provided
would often constitute visually poor fits.

To combat this, an iterative, self-consistent approach was employed. After
computing the estimates described above, using all available data points, the
resulting values were then used to calculate the limits. Data points would then
be removed from the fits based on these new limits and the process repeated and
the limits re-calculated. Data points would then be removed again (or at this
point possibly re-added) based on the new limit values. This would repeat until
no further changes occurred --- that is the results became self-consistent. In
some cases, where such a method would lead to all points being removed in early
iterations (causing the process to terminate prematurely), a more simplistic
approach of removing low-temperature points one by one would be used until the
estimates became self-consistent.

After this, fits would normally still appear visually to be sub-optimal, thus a
final, generalised fitting would be run. In this, the whole expression would be
fit to, using a least-squares, Nelder-Mead optimisation function. The previously
estimated values would be used as the initial values for the fit, with their
values being constrained to remain within 30\% of these values.

\section{Results}

By performing the self-consistent analysis presented in the previous section,
three sets of Hall data from
Figures~\ref{fig:data:achall}~and~\ref{fig:data:dchall} were fitted (with
maximum conductivities: $\sim$\,600\,S/cm, $\sim$\,750\,S/cm,
$\sim$\,1000\,S/cm). The other two sets of data proved to have too few points
and too much noise to reliably fit. The results of these fittings are shown in
Figure~\ref{fig:mainTable}.

Focusing first on the values of $T_*$, it should be possible to glean some
insights. In our chosen function for $b(T)$, $T_*$ controls the temperature
above which the ratio of non-deflectable to deflectable mobilities plateaus and
becomes relatively temperature independent. Therefore, a greater value of $T_*$
would correspond to a system that requires a greater value of temperature to
achieve relatively temperature-independent transport. Therefore, it could be
said that a greater value of $T_*$ indicates a greater amount of energetic
disorder, with its transport remaining thermally activated for greater values of
temperature. Following this line of logic, all three devices have similar, small
values of $T_*$, with a slight decrease being apparent at the greatest
conductivity, albeit within the error ranges of the other two devices. The small
values of $T_*$ here (relative to room temperature) indicate relatively low
levels of disorder overall. If the decrease in device 3 can be trusted, it would
suggest that the energetic disorder of such systems continues to decrease with
increased doping. This is all consistent with our recent study of the
charge-transport physics of IEx-doped conjugated polymers, including
PBTTT,\cite{Jacobs2021, Jacobs2022} where grazing-incidence wide-angle x-ray
scattering (GIWAXS) measurements of IEx-doped PBTTT films showed reduced $\pi -
\pi$ stacking disorder at high doping concentrations. There is also evidence
that the intrachain and interchain polaron delocalisation lengths become very
long at high doping concentrations allowing the charges to average effectively
over individual Coulomb wells and reducing the energetic disorder felt by the
carriers. This would also appear to be supported by the value of $1/(1 + h)$
increasing slightly with doping concentration, from 15.4\% to 16.4\%, which would
indicate a greater degree of deflectable transport contributing to conduction.
However, the errors on these values overlap, hence it is hard to say with any
certainty if this is truly what is happening in these data, or whether this is
simply due to random error.

The maximum $\av{\mu_d}$ values extracted here range around
$\mu$\,$\sim$\,4\,--\,8\,cm$^2/$Vs. This is consistent with mobilities estimated
from carrier densities measured by XPS previously (6\,--\,8\,cm$^2/$Vs) and an order
of magnitude greater than typical values for undoped PBTTT in FET
architectures.\cite{McCulloch2006} This similarly supports the idea that these
doped films exhibit significantly less energetic disorder as a result of their
doping. Furthermore, the two most highly doped devices here exhibit a
significantly less temperature-dependent deflectable mobility than the least
doped device. This similarly suggests that these devices exhibit a greater
amount of order, with deflectable transport becoming more band-like in its
temperature dependence at greater doping levels. This finding has important
implications: it shows that a new regime has been reached, where not only does
doping to high concentrations not disrupt the crystalline and energetic order of
the polymer, but goes so far as to enhance it. In previous work on
solid-state-diffusion-doped PBTTT,\cite{Kang2016} where only the uncorrected
values of Hall mobility were available for analysis (with those values being
underestimates and effectively only acting as a lower bound on the true value)
it was only possible to conclude that doping did not significantly disrupt the
lattice structure of the polymer.\cite{Kang2016} Now that more meaningful values
of mobility can be extracted, as well as $T_*$ and $1/(1 + h)$, it is clear that
this may have been something of an understatement.

The values of $\av{n_d}$ for all three devices are much more physically
reasonable values than those extracted by simply assuming $R_H = [q n]^{-1}$ in
Figures~\ref{fig:data:acdensity}~and~\ref{fig:data:dcdensity}. By assuming a
monomer density of $\sim 1 \times 10^{21}$\,cm$^{-3}$, estimated from the
polymer film density, these suggest that devices 1, 2 and 3 (on average) have
roughly one deflectable polaron per 9, 8, and 3 monomer sites respectively.

Further to this, the total charge-carrier density of a PBTTT device, similarly
doped to device 3 here, was measured in our previously mentioned, recent
charge-transport study\cite{Jacobs2022} using x-ray photoemission spectroscopy
(XPS) and nuclear magnetic resonance (NMR). By determining the concentration of
TFSI$^-$ anions in the film, its total charge-carrier density was determined to
be $n_t = (8.84 \pm 0.24) \times 10^{20}$\,cm$^{-3}$. Since $\av{n_d} = \av{g}
n_t$, this would suggest that device 3 has a value of $\av{g} \approx 0.37$.
That is, carriers spend $\sim$\,37\% of their time moving deflectably. Following
from this, the result here of deflectable transport only accounting for
$\sim$\,16\% of the conductivity despite carriers being in such states for
$\sim$\,37\% of the time, suggests that the carriers undergoing less deflectable
transport may in fact exhibit higher carrier mobilities than those in more
deflectable pathways. This is not necessarily a contradiction, as more
deflectable pathways are likely to consist of more highly ordered, crystalline
polymer domains that support more delocalised carriers, but may not be well
interconnected. This also suggests that the limiting factor on conductivity in
such deflectable pathways is mobility rather than charge-carrier density. This
further solidifies the observation in our previous study that the limit of
conductivity enhancement that can be achieved by adding charge carriers has been
reached with IEx doping. Indeed, the fact that the most highly doped device here
(device 3) has a significantly smaller maximum deflectable mobility than the
other two, lesser-doped devices suggests that something of a trade-off is
occurring between carrier density and mobility --- where greater carrier
densities provide diminishing returns on their enhancement of conductivity.

\section{Conclusions}

We have presented a new methodology for interpreting Hall effect measurements on
highly doped PBTTT, based on analysing temperature-dependent Hall data. To
provide said data, we have developed an improved AC Hall measurement system and
data-extraction routine, allowing measurements of the Hall effect from 10K to
room temperature with high signal-to-noise ratios. Using PBTTT as a model system
we have shown that the improved Hall analysis is able to determine the
``deflectable'' carrier concentration and mobility. Not only that, but our
method also provides insight into how ``ideal'' the transport of charge carriers
is by allowing us to estimate the relative contributions to the total
conductivity from more ordered regions of the polymer, where carriers can be
deflected by magnetic fields, and from less ordered regions where they cannot.
Our method and analysis promises to transform the Hall effect from a finicky and
hard-to-interpret transport phenomenon in polymer semiconductors, into a
valuable tool for understanding and optimising such materials in the future.

\begin{acknowledgements}
    Financial support from the European Research Council for a Synergy grant
    SC2 (no.~610115) and from the Engineering and Physical Sciences Research
    Council (EP/R031894/1) is gratefully acknowledged.  W.A.W. acknowledges
    funding through the EPSRC Integrated Photonic and Electronic Systems CDT
    (now known as the Connected Electronic and Photonic Systems CDT). H.S.
    acknowledges support from a Royal Society Research Professorship
    (RP\textbackslash R1\textbackslash 201082). Furthermore, W.A.W. would
    like to thank both Roger Beadle and Thomas Sharp, technicians within the
    Cavendish Laboratory, for their help with designing and building the
    experimental system detailed in this article.
\end{acknowledgements}

\end{document}